# Kadomtsev-Petviashvili II equation: Structure of asymptotic soliton webs

Shai Horowitz
Department of mathematics
Rutgers University
Piscataway, NJ 08854-8019, USA

and

Yair Zarmi
Jacob Blaustein Institutes for Desert Research
Ben-Gurion University of the Negev
Midreshet Ben-Gurion, 8499000 Israel

Abstract

A wealth of observations, recently supported by rigorous analysis, indicate that, asymptotically in time, most multi-soliton solutions of the Kadomtsev-Petviashvili II equation self-organize in webs comprised of solitons and soliton-junctions. Junctions are connected in pairs, each pair - by a single soliton. The webs expand in time. As distances between junctions grow, the memory of the structure of junctions in a connected pair ceases to affect the structure of either junction. As a result, every junction propagates at a constant velocity, which is determined by the wave numbers that go into its construction. One immediate consequence of this characteristic is that asymptotic webs preserve their morphology as they expand in time. Another consequence, based on simple geometric considerations, explains why, except in special cases, only 3-junctions ("*Y*-shaped", involving three wave numbers) and 4-junctions ("*X*-shaped", involving four wave numbers) can partake in the construction of an asymptotic soliton web.

Key words: KP II equation; soliton webs; junction "lattice"

PACS: 05.45.Yv, 05.90.+m

## 1. Introduction

The construction and properties of soliton solutions of the Kadomtsev-Petviashvili II equation (KP II) [1],

$$\frac{\partial}{\partial x}\left(-4\frac{\partial u}{\partial t} + \frac{\partial^3 u}{\partial x^3} + 6u\frac{\partial u}{\partial x}\right) + 3\frac{\partial^2 u}{\partial y^2} = 0 \quad , \tag{1}$$

have been studied extensively in the literature [2-39]. Solution characteristics that emerge from the literature and are relevant to the subject matter of this paper are reviewed in the following.

All solutions constructed from three wave numbers, called Miles resonances [2-4] or *Y*-shaped solutions [21,24-35], display one junction (see Fig. 1). A single-junction solution, constructed from four wave numbers, the *X*-shaped solution, has been also identified [5, 24-35] (see Fig. 2). (In the figures, pairs, $\{i,j\}$, of integers near a soliton represent the wave numbers, $\{k_i, k_j\}$, from which it is constructed.)

Except for the *X*-shaped solution, a salient feature of multi-soliton solutions that are constructed from $M \geq 4$ wave numbers is their self-organization, as $t \rightarrow \pm\infty$, into webs comprised of soliton-lines and soliton-junctions [24-39]. Some solitons emanate from junctions and are not connected to other junctions, whereas others connect junctions. At finite times, a solution, which asymptotically self-organizes into a web, may exhibit a short-lived structure, e.g., a single junction, or a more complex structure that covers a finite domain in the *x*-*y* plane. However, as $t \rightarrow \pm\infty$, the short-lived structure evolves into the asymptotic web. Asymptotically persistent junctions are connected in pairs, each pair by one soliton. Each junction propagates in the *x*-*y* plane at a constant velocity, which is determined by the wave numbers that go into its construction. The $t \rightarrow +\infty$ and $t \rightarrow -\infty$ asymptotic webs need not be identical.

The (4,1) solution offers an example for these statements. For some finite duration, it may exhibit a single junction, from which four soliton lines emerge. In time, it develops two 3-junctions, which are constructed out of three wave numbers and contain three solitons ("*Y*-shaped"; see Fig. 3). The junctions become infinitely removed from one another as $|t| \to \infty$. They are connected by a single soliton. The junctions in the $t \to +\infty$ and $t \to -\infty$ webs are constructed form different wave solitons, but the morphology of the webs is the same in both limits [24-39].

Another example is the family of (4,2) solutions. In most solutions in this family, a single junction exists only for a finite duration and evolves into several junctions as $t \to \pm\infty$. The most common case is that of a web that contains four 3-junctions (see Fig. 4); the same junctions emerge at both asymptotic limits [24-35]. Specific (4,2) solutions may exhibit a single 4-junction ("*X*-shaped", with four emerging solitons, constructed out of four wave numbers) at either $t \to +\infty$ or $t \to -\infty$, accompanied by 3-junctions. (An example will be provided in Section 4.) The *X*-shaped solution represents a unique case. It is a (4,2) solution that displays only one 4-junction, which persists both as $t \to +\infty$ and $t \to -\infty$ [5, 24-35].

Finally, the analysis of many solutions [24-39] points at two empirical rules:

*(i)* The webs expand in time, but preserve their morphology;

*(ii)* Except for special wave-number combinations, only 3- and 4-junctions partake in the construction of asymptotic webs.

Rigorous demonstrations of the self-organization of solutions into asymptotic webs have existed in the case of (*M*,1) solutions [36-38], for all $M \geq 3$, and of the (4,2) family of solutions [28]. As for (*M*,*N*) solutions with $M \geq 4$, $N > 2$, the observations reviewed above have been based on attempts at classifying solution patterns as well as on the rigorous or numerical study of many solutions [24-35]. The situation has changed with the appearance of Refs. [40-43], where the Sato representation of soliton solutions of the KP II equation in terms of points of the totally non-negative part of the Grassmannian $(Gr_{k,n}) \geq 0$ [9] is analyzed with tools of Combinatorics and Graph Theory.

Refs. [40-43] provide a rigorous and detailed classification of an important class of regular solutions of the KP II equation and of the higher members of its hierarchy - the class of solutions that are constructed out of "generic" wave-number sets (defined in Section 2.7). The observations reviewed above are amongst the results of this analysis.

The goal of this paper is of a limited scope: to show that rules (*i*) and (*ii*) can be understood with the aid of simple geometric considerations that are direct consequences of the construction algorithm of the solutions, reviewed in Section 2. Specifically, all information regarding the structure of a junction is lost at an exponential rate along each soliton line that emanates from the junction. As a result, asymptotically removed junctions do not affect one another. Each junction tends to a separate single-junction solution. The latter propagates at a constant velocity, which is determined by the wave numbers that go into its construction. This leads to rule (*i*): As $|t| \to \infty$, the soliton web expands linearly in time in both the *x*- and *y*-directions. Hence, the morphology of the web is not changed as it expands. This characteristic has been the basis for the notion of asymptotic contour plots [40-43]. The main point of this paper is that, based on the web characteristics reviewed above, a simple geometric consideration also explains rule (*ii*).

The construction of soliton solutions of Eq. (1) and their properties that are pertinent for the points raised in this paper are reviewed in Section 2 and Appendix A. Section 3 presents the working hypothesis and the geometric considerations, which explain why expanding junctions preserve their morphology and why an asymptotic soliton web can contain only 3- and 4-junctions. Section 4 presents several numerical examples.

## 2. Review of KP II line-soliton solutions
### 2.1 Construction of solutions
The line-soliton solutions of Eq. (1) are obtained through the transformation [14, 15, 23, 26-28]:

$$u(t,x,y) = 2\partial_x^2 \log\{f(t,x,y)\} \equiv u(M,N;\vec{\xi}) \quad . \tag{2}$$

For a pre-selected set of $M$ wave numbers, $\{k_1,\ldots,k_M\}$, the function $f(t,x,y)$ is given by

$$f(t,x,y) \equiv f(M,N;\vec{\xi}) =$$

$$\begin{cases} \sum_{i=1}^{M} \xi_M(i) \exp(\theta_i(t,x,y)) & N=1 \\ \sum_{i=1}^{M} \xi_M(i) \exp\left(\sum_{j=1, j\neq i}^{M} \theta_j(t,x,y)\right) & N=M-1 \\ \sum_{1\leq i_1<\ldots<i_N\leq M} \xi_M(i_1,\ldots,i_N) \left(\prod_{1\leq j<l\leq N}(k_{i_l}-k_{i_j})\right) \exp\left(\sum_{j=1}^{N}\theta_{i_j}(t,x,y)\right) & 2\leq N\leq M-2 \end{cases}, \quad (3)$$

$$k_1 < k_2 < \ldots < k_M \;, \tag{4}$$

$$\theta_i(t,x,y) = -k_i x + k_i^2 y - k_i^3 t \;. \tag{5}$$

The sums in Eq. (3) go over all of $\binom{M}{N}$ subsets of $1 < N < M-1$ wave numbers in $\{k_1,\ldots,k_M\}$.

A solution, $u(M, N; \vec{\xi})$, computed at $(-t, -x, -y)$, coincides with another solution, $u(M, M-N; \vec{\xi})$ computed at $(t,x,y)$ [23-36]. Hence, it suffices to study solutions with $N \leq [M/2]$. The $\xi$-coefficients are called the Plücker coordinates.

Finally, the construction of the solution through Eqs. (2)-(5) ensures that information regarding a soliton junction in an asymptotic web decays exponentially fast along each soliton that emanates from the junction. Away from the junction, the soliton does not transmit any information regarding the structure of the junction. Examples to this effect will be presented in Sections 2.6 and 2.8.

**2.2 Constraints on Plücker coordinates**
To exclude singular solutions, one requires non-negativity of all Plücker coordinates:

$$\xi_M(i_1,\ldots,i_N) \geq 0 \;. \tag{6}$$

For $N = 1$ and $N = M-1$, $\xi_M(i)$ are arbitrary constants that obey Eq. (6). For $2 \leq N \leq M-2$, $\xi_M(i_1,\ldots,i_N)$ are constrained also by the Plücker relations (see, e.g., [23]). For example, a single Plücker relation exists in the case of an $(M,N) = (4,2)$ solution:

$$\xi_4(1,2)\xi_4(3,4) - \xi_4(1,3)\xi_4(2,4) + \xi_4(1,4)\xi_4(2,3) = 0 \quad . \tag{7}$$

A way to obtain these relations is presented in Appendix A.

### 2.3 Single-soliton solution
The single-soliton solution of Eq. (3) is:

$$u(t,x,y) = \frac{2\left((k_2 - k_1)/2\right)^2}{\left(\cosh\left[\left((k_2 - k_1)/2\right)\left(x - (k_1 + k_2)y + (k_1^2 + k_1 k_2 + k_2^2)t\right) + \ln\left(\sqrt{\xi_1/\xi_2}\right)\right]\right)^2} \quad . \tag{8}$$

The slope of a soliton line in the *x-y* plane, previously exploited in the analysis of multi-soliton solutions of Eq. (1) in refs. [24-28, 30], will be exploited in Section 3. It is equal to:

$$\frac{dy}{dx} = \frac{1}{k_1 + k_2} \quad . \tag{9}$$

### 2.4 "Riding" on soliton junction: *Y*-shaped solution
In the context of the KP II equation, the idea of "riding" on soliton junctions, namely, of viewing a solution in a frame of reference that is moving at the velocity of a junction, so that the latter is stationary, was first employed in the search for rational solutions of the equation, which propagate at a constant velocity [20]. It was then used in the study of solutions with $M = 3$ in Eq. (3) [21]. The latter all display one 3-junction; they are the Miles resonances [2-4] or *Y*-shaped solutions [24-35] (see Fig. 1). Eq. (3) then has the form:

$$f(t,x,y) = \xi_3(1)e^{\theta(t,x,y;k_1)} + \xi_3(2)e^{\theta(t,x,y;k_2)} + \xi_3(3)e^{\theta(t,x,y;k_3)} \tag{10}$$

The solution moves rigidly at a constant velocity. In a frame of reference moving at that velocity, the solution is stationary [21]:

$$f(t, x + v_{3,x}t, y + v_{3,y}t) = e^{\sigma t}f(0,x,y) \quad \Rightarrow \quad u(t, x + v_{3,x}t, y + v_{3,y}t) = u(t=0,x,y) \quad , \tag{11}$$

$$\sigma = -k_1 k_2 k_3 \quad , \quad v_{3,x}(k_1,k_2,k_3) = k_1 k_2 + k_1 k_3 + k_2 k_3 \quad , \quad v_{3,y}(k_1,k_2,k_3) = k_1 + k_2 + k_3 \quad . \tag{12}$$

The subscript "3" in Eqs. (11) and (12) signifies that $\vec{v}_3$ is constructed out of three wave numbers.

## 2.5 "Riding" on soliton junction: *X*-shaped solution

The *X*-shaped solution [5, 24-35] is a well-known single-junction solution. (See Fig.2.) It is a (4,2) solution, which exists only for each of the following choices of the $\xi$- coefficients:

$$\xi_4(1,2) = \xi_4(3,4) = 0 \quad , \quad \xi_4(1,3) = \xi_4(2,4) = 0 \quad , \quad \xi_4(1,4) = \xi_4(2,3) = 0 \quad . \tag{13}$$

For each choice in Eq. (13), the solution propagates at a constant velocity, preserving its shape. Eq. (11) is obeyed for each of the choices. The values of the velocity and of $\sigma$ in Eq. (12) are different for each choice. A plot of the *X*-shaped solution corresponding to the first choice in Eq. (13) is shown in Fig. 2. Its velocity is given by

$$\begin{aligned} v_{4,x}(k_1,k_2,k_3,k_4) &= k_1 k_2 + k_1 k_3 + k_1 k_4 + k_2 k_3 + k_2 k_4 + k_3 k_4 - \frac{k_3 k_4 (k_3 + k_4) - k_1 k_2 (k_1 + k_2)}{(k_3 + k_4) - (k_1 + k_2)} \\ v_{4,y}(k_1,k_2,k_3,k_4) &= k_1 + k_2 + k_3 + k_4 - \frac{k_3 k_4 - k_1 k_2}{(k_3 + k_4) - (k_1 + k_2)} \end{aligned} \tag{14}$$

The subscript "4" signifies that $\vec{v}_4$ is constructed out of four wave numbers. The 4-velocities corresponding to the second and third possibilities in Eq. (13) are simple modifications of Eq. (14).

## 2.6 "Riding" on soliton junctions: (4,1) and (4,2) solutions

There are two types of solutions with *M* = 4 wave numbers: (4,1) and (4,2) solutions. With four wave numbers, one can form four triplets, hence four *prospective* 3-junctions may exist: {1,2,3}, {1,3,4}, {1,2,4} and {2,3,4}. ({*i,j,l*} indicates the wave numbers, {$k_i$, $k_j$, $k_l$}, from which a 3-junction is constructed.) Which of them persist asymptotically in time, depends on the solution.

(4,1) solution The asymptotic web of this solution contains four external solitons and two 3-junctions that are connected by one soliton (see Fig. 3). In a frame of reference that moves at the 3-velocity (see Eq. (12)) of either the {1,2,3}- or the {1,3,4}-junction, the corresponding junction is at rest. One then finds:

$$\begin{aligned}
u\bigl(t, x + v_{3,x}(k_1,k_2,k_3)t, y + v_{3,y}(k_1,k_2,k_3)\bigr) &\underset{t\to+\infty}{\to} Y(x,y;1,2,3) \\
u\bigl(t, x + v_{3,x}(k_1,k_3,k_4)t, y + v_{3,y}(k_1,k_3,k_4)\bigr) &\underset{t\to+\infty}{\to} Y(x,y;1,3,4) \\
u\bigl(t, x + v_x(k_1,k_2,k_3)t, y + v_{3,y}(k_1,k_2,k_3)\bigr) &\underset{t\to-\infty}{\to} 0 \\
u\bigl(t, x + v_{3,x}(k_1,k_3,k_4)t, y + v_{3,y}(k_1,k_3,k_4)\bigr) &\underset{t\to-\infty}{\to} 0
\end{aligned} \quad (15)$$

The two 3-junctions persist as $t \to +\infty$: The solution tends to $Y(x,y;i,j,l)$, the corresponding Y-shaped solution (defined by Eqs. (11)-(13)), *at rest*. Although connected by a {1,3} soliton, the two infinitely removed junctions do not affect one another. In the same two frames of reference, the (4,1)-solution tends to zero as $t \to -\infty$. The two junctions do not persist. Similarly, viewing the solution in a frame of reference that moves at the velocities of either of the two remaining 3-junctions ({1,2,4} and {2,3,4}) one finds that they persist as $t \to -\infty$, but not as $t \to +\infty$, and each is unaffected by the other.

(4,2) solution When all $\xi_4(i,j) \neq 0$ in Eqs. (2)-(5), the asymptotic web has the form shown in Fig. 4. The "riding on junction" analysis yields that all four junctions persist at both $t \to \pm\infty$, each unaffected by the junctions to which it is connected.

**2.7 Generic wave number sets**
The multi-soliton solutions of Eq. (1) have been classified for the large subset of regular solutions (obeying Eq.(6)) with generic wave-number sets [40-43]. Define the following sum:

$$S_M^N(i_1,i_2,\ldots,i_N) \equiv \sum_{j=1}^N k_{i_j} \quad , \quad k_{i_1} < k_{i_2} < \ldots < k_{i_N} \quad , \quad N < M \quad . \quad (16)$$

The set of $M$ wave numbers, from which the solution is constructed is called *generic* if all $S_M^N(i_1,i_2,\ldots,i_N)$ are distinct; they do not obey resonance relations. Namely, for any $N \neq N'$

$$S_M^N(i_1,i_2,\ldots,i_N) \neq S_M^{N'}(j_1,j_2,\ldots,j_{N'}) \quad . \quad (17)$$

This constraint ensures, in particular, that the velocities of all junctions in an asymptotic web are all different from one another.

If, instead of the inequality, equality holds in Eq. (17), then the solutions *may* be degenerate. The 4-junction (*X*-shaped) solution provides a simple example. When the denominator in Eq. (14) vanishes:

$$(k_3 + k_4) = (k_1 + k_2) \; , \qquad (18)$$

the velocity of the 4-junction becomes infinite; the solution degenerates into two parallel single solitons. However, the structure of a (4,2) solution, in which all $\xi_4$ coefficients are nonzero, as in Fig. 4, is not affected when a non-generic set of four wave numbers is used.

**2.8 Loss of "memory" away from junction and factorization of solutions**
The mere construction algorithm of the solution through Eqs. (2)-(5) ensures that, away from a junction all memory of the structure of the junction is lost exponentially fast along a soliton line that emanates from the junction. The (4,1) and (4,2) solutions, discussed in Section 2.6, provide an example to this statement. Another way to see this is through mapping of a multi-soliton solution with the aid of the differential polynomial

$$R[u] = u^3 + u u_{xx} - (u_x)^2 \; . \qquad (19)$$

The properties of *R*[*u*] have been discussed in detail in Ref. [44], hence they are briefly summarized here. *R*[*u*] vanishes identically on any single-soliton solution of the KP II equation, and maps the asymptotic web of a multi-soliton solution into a collection of "vertices" - positive definite humps that are localized around the soliton junctions. Along each soliton line, *R*[*u*] falls off exponentially as a function of the distance, *d*, from the junction. For example, for the 3-junction (*Y*-shaped) solution, described in Section 2.4, at any fixed time *R*[$u_Y$] falls off as:

$$R[u_Y] \propto e^{-ad} \; . \qquad (20)$$

In Eq. (20), *a* is a known coefficient. The analysis yields similar results for the 4-junction (*X*-shaped) solution, described in Section 2.5.

This observation is the basis of the notion of the "factorization" of the asymptotic web, examples for which have been discussed in Section 2.6: If one "rides" on a junction, i.e., one views the solution in a frame that moves at the velocity of that junction, the solution tends to a single-junction solution of the KP II equation, and all other junctions are removed to infinity in the *x-y* plane.

**3. Structure of asymptotic webs**
**3.1 Definitions, notation, and working hypothesis**
<u>Definition</u> A *p*-junction is a junction, at which *p* soliton lines intersect, constructed out of a set of $p \geq 3$ wave numbers, $\{k_{i_1},...,k_{i_p}\}$. It will be denoted by $J_p\{i_1,...,i_p\}$. For example, a *Y*-shaped junction is a 3-junction and an *X*-shaped junction is a 4-junction.

<u>Definition</u> A *p*-velocity is the velocity vector of a *p*-junction, denoted by $\vec{v}_p\{k_{i_1},...,k_{i_p}\}$:

$$\vec{v}_p\{k_{i_1},...,k_{i_p}\} = \{v_{p,x}\{k_{i_1},...,k_{i_p}\}, v_{p,y}\{k_{i_1},...,k_{i_p}\}\} \ . \tag{21}$$

For example, Eqs. (12) and (14) represent a 3- and a 4-velocity, respectively. For a set of $p \geq 4$ wave numbers, there may be several *p*-junctions, with different *p*-velocities (see Section 3.2).

<u>Notation</u> A product of *N* exponentials appearing in Eq. (3) will be denoted by

$$\pi\{i_1,...,i_N\} \equiv \exp\left(\sum_{j=1}^{N} \theta_{i_j}(t,x,y)\right) \ . \tag{22}$$

The construction of the solutions of Eq. (1) through Eqs. (2)-(5) ensures that the effect of a junction decays exponentially fast along any soliton line, which emanates from the junction. A soliton that connects two remote junctions cannot convey information between the two except for its wave-number content. This is the basis for the following working hypothesis:

<u>Working hypothesis</u> *As $t \rightarrow \pm\infty$, the web contains junctions that are connected in pairs. Each pair is connected by a single soliton. As the distance between any two connected junctions becomes*

*infinite, the junctions do not affect one another and move each at a constant velocity that is determined by the wave numbers from which it is constructed.*

### 3.2 Prospective *p*-junctions

To find *p*-velocities of prospective *p*-junctions in an (*M,N*) solution, one views the solution in a frame that is moving at the sought-for velocity. The phase of a term, $\pi\{i_1,\ldots,i_N\}$ (see Eq. (22)), in the moving frame is given by

$$\sum_{j=1}^{N}\theta_{i_j}\left(t, x+v_x t, y+v_y t\right) = \left(\sum_{j=1}^{N}\left(-k_{i_j}^3 - k_{i_j} v_x + k_{i_j}^2 v_y\right)\right)t + \left(\sum_{j=1}^{N}\left(-k_{i_j}\right)\right)x + \left(\sum_{j=1}^{N}\left(k_{i_j}^2\right)\right)y \ . \quad (23)$$

As the velocity has two components, every triplet of $\pi$-terms in the function *f(t,x,y)* in Eq. (3) has the capacity to determine a velocity, which guarantees that the time dependence of the three members of the triplet coincides. Selecting a triplet of $\pi$-terms, one requires:

$$\left(\sum_{j=1}^{N}\left(-k_{i_j}^3 - k_{i_j} v_x + k_{i_j}^2 v_y\right)\right) = \left(\sum_{m=1}^{N}\left(-k_{i_m}^3 - k_{i_m} v_x + k_{i_m}^2 v_y\right)\right) = \left(\sum_{n=1}^{N}\left(-k_{i_n}^3 - k_{i_n} v_x + k_{i_n}^2 v_y\right)\right) \ . \quad (24)$$

In Eq. (24), the indices (*j,m,n*) represent three *different* subsets of *N* wave numbers appearing in the sum in Eq. (3). Each such triplet determines one prospective velocity vector. Different triplets may determine different or identical velocities.

There is an infinity of solutions for the *p*-velocities: For $3 \leq M < \infty$ and $1 \leq N \leq [M/2]$, Eq. (24) allows for *p*-velocities with $3 \leq p < \infty$. A solution for a *p*-velocity exists when the following two conditions are obeyed for an (*M,N*) solution:

$$M \geq p \ , \quad 3N \geq p \ . \quad (25)$$

The following are examples of possible solutions of Eq. (24) for some (*M,N*) solutions: Only 3-velocities in the case of *any* (*M*,1) solution; 3- and 4-velocities in the case of a (4,2) solution; and 3-, 4- and 5-velocities in the case of a (5,2) solution. A 6-velocity first solves Eq. (24) in the cases of (6,2) and (6,3) solutions. 3-, 4-, 5- and 6-velcoities may solve Eq. (24) in the case of (*M*,2) solutions, with $M \geq 4$. Finally, *p*-velocities with $p > 6$, exist only when both $M > 6$ and $N > 2$.

Finding *p*-velocities is only the first step. One must then find whether these velocities correspond to *p*-junctions that exist and persist as $t \to \pm\infty$, so as to partake in the construction of the asymptotic web. For such junctions to exist and persist, the function $f(t,x,y)$ of Eq. (3) must contain terms that generate the junctions, and these terms must dominate as $t \to +\infty$ and/or as $t \to -\infty$, when viewed in a frame that moves at the corresponding *p*-velocity. The detailed analysis leading to identification of asymptotically persistent junctions is left for a separate publication. In the following, as an example, it is shown that 5- and 6-velocities cannot correspond to 5- and 6-junctionc that partake in the construction of an asymptotic web.

### 3.3 Why only 3- and 4-junctions?

As $t \to \pm\infty$, the line connecting two junctions in the *x-y* plane tends to a single soliton. There are then two expressions for the slope of that line. The first expression is the slope of the single-soliton line, given in Eq. (9). The second is determined by the velocity vectors of the two junctions. Denoting the wave numbers of the connecting soliton by $k_1$ and $k_2$, these two must partake in the construction of both junctions. Other than that, each junction may be constructed from different wave numbers. One now requires that the two expressions for the slope coincide:

$$\frac{v_{p,y}\{k_1,k_2,k_3,...,k_p\} - v_{p',y}\{k_1,k_2,k'_3,...,k'_{p'}\}}{v_{p,x}\{k_1,k_2,k_3,...,k_{p_1}\} - v_{p',x}\{k_1,k_2,k'_3,...,k'_{p'}\}} = \frac{1}{k_1+k_2} \quad . \tag{26}$$

In Eq. (26), the prime indicates that, except for $k_1$ and $k_2$, the wave number sets in the two junctions, may differ in size (*p* and *p'*) as well as in the numerical values of the wave numbers. Let us now recast Eq. (26) as

$$\begin{aligned} v_{p,x}\{k_1,k_2,k_3,...,k_p\} - (k_1+k_2)v_{p,y}\{k_1,k_2,k_3,...,k_p\} = \\ v_{p',x}\{k_1,k_2,k'_3,...,k'_{p'}\} - (k_1+k_2)v_{p',y}\{k_1,k_2,k'_3,...,k'_{p'}\} \end{aligned} \quad . \tag{27}$$

The following discussion applies to the non-degenerate case, namely, that all wave numbers may be treated as independent entities. The constraint that the set of $M$ wave numbers in Eq. (4) is generic [40-43] (see Section 2.7) is a sufficient condition for this requirement. As the two junctions do not affect one another when the distance between them tends to infinity, neither the l.h.s. nor the r.h.s. of Eq. (27) can depend on any wave numbers except for $k_1$ and $k_2$:

$$v_{p,x}\{k_1,k_2,k_3,...,k_p\} - (k_1 + k_2)v_{p,y}\{k_1,k_2,k_3,...,k_p\} = \\ v_{p',x}\{k_1,k_2,k_3,...,k_{p'}\} - (k_1 + k_2)v_{p',y}\{k_1,k_2,k'_3,...,k'_{p'}\} = g(k_1,k_2) \quad . \tag{28}$$

Next, consider a connection of one of the junctions, say, the $p$-junction, with a third junction. There are three possibilities for the soliton that connects the two.

<u>Soliton with same wave numbers, $\{k_1, k_2\}$</u> This is possible when the same soliton emanates from the $p$-junction in two directions, as in the $X$-shaped solution, (see Fig. 2). Eq. (28) is generated again for the $p$-junction, hence, no new constraint is obtained.

However, as a junction must contain at least one additional, different, soliton, the following two cases yield constraints of significance.

<u>Soliton with two different wave numbers, $\{k_3, k_4\}$</u> Eq. (28) for the additional connection reads:

$$v_{p,x}\{k_1,k_2,k_3,k_4,...,k_p\} - (k_3 + k_4)v_{p,y}\{k_1,k_2,k_3,k_4,...,k_p\} = h(k_3,k_4) \quad . \tag{29}$$

Consistency of Eqs. (28) and (29) requires that the velocity under consideration must be constructed out of four wave numbers only; it must represent a 4-juncion, $J_4\{1,2,3,4\}$.

<u>Soliton with one different wave number, e.g., $\{k_1, k_3\}$</u> Eq. (28) for the additional connection reads:

$$v_{p,x}\{k_1,k_2,k_3,...,k_p\} - (k_1 + k_3)v_{p,y}\{k_1,k_2,k_3,...,k_p\} = \tilde{h}(k_1,k_3) \quad . \tag{30}$$

Consistency of Eqs. (28) and (30) requires that the velocity under consideration must be constructed out of three wave numbers only; it must represent a 3-juncion, $J_3\{1,2,3\}$.

In summary, only 3- and 4-junctions may appear in a soliton web as $t \to \pm\infty$. For these, the unspecified functions on the r.h.s of Eqs. (28)-(30) obtain a simple expression. The velocity vector of the 3-junction, given by Eq. (12), obeys Eq. (28) for all three pairs of wave numbers:

$$\begin{aligned}v_{3,x}\{k_1,k_2,k_3\}-(k_1+k_2)v_{3,y}\{k_1,k_2,k_3\}&=-k_1^2-k_1k_2-k_2^2\\v_{3,x}\{k_1,k_2,k_3\}-(k_1+k_3)v_{3,y}\{k_1,k_2,k_3\}&=-k_1^2-k_1k_3-k_3^2\\v_{3,x}\{k_1,k_2,k_3\}-(k_2+k_3)v_{3,y}\{k_1,k_2,k_3\}&=-k_2^2-k_2k_3-k_3^2\end{aligned} \quad (31)$$

(Eq. (31) was exploited in Ref. [21] as an intermediate step in obtaining Eq. (12) from Eq. (11)).

The velocity vector of the 4-junction, given by Eq. (14), obeys both Eqs. (28) and (29):

$$\begin{aligned}v_{4,x}\{k_1,k_2,k_3,k_4\}-(k_1+k_2)v_{4,y}\{k_1,k_2,k_3,k_4\}&=-k_1^2-k_1k_2-k_2^2\\v_{4,x}\{k_1,k_2,k_3,k_4\}-(k_3+k_4)v_{4,y}\{k_1,k_2,k_3,k_4\}&=-k_3^2-k_3k_4-k_4^2\end{aligned} \quad (32)$$

The significance of Eqs. (31) and (32) is that each soliton line that emanates from either junction can connect it to any other junction that shares the same soliton.

A final comment is due at this point. If all the wave numbers cannot be viewed as independent entities, in particular, if the set of $M$ wave numbers, from which the solution is constructed, is not generic (see, Section 2.7), then $p$-velocities with $p > 4$ may solve Eqs. (28) and/or (29). Such $p$-velocities may correspond to $p$-junctions that do partake in the construction of asymptotic webs of solutions of higher members of the KP II hierarchy [40-43], which are not discussed here.

### 3.4 Persistent junctions with $p = 5$ & 6 wave numbers?

As argued in Section 3.2, Eq. (24) allows for $p$-velocities with $p > 4$. However, the analysis in Section 3.3 precludes the possibility that such velocities can correspond to junctions that partake in the construction of an asymptotic web. As an example to this statement, the cases of 5- and 6-velocities, are discussed through the analysis of $(M,2)$ solutions with $M \geq 5$.

In the notation of Eq. (22), each term in $f(t,x,y)$ of Eq. (3) is then a product, $\pi\{i_1, i_2\}$, of two exponentials. The highest value allowed by Eqs. (23) and (24) is $p = 6$. Solving Eq. (24) for a triplet

($\{\pi\{1,2\},\pi\{3,4\},\pi\{5,6\}\}$), with six different wave numbers, yields the expression for the 6-velocity vector:

$$v_{6,x}\{k_1,k_2,k_3,k_4,k_5,k_6\} = \left\{\begin{matrix}(k_1^2+k_2^2)(k_5^3+k_6^3-k_3^3-k_4^3)\\+(k_3^2+k_4^2)(k_1^3+k_2^3-k_5^3-k_6^3)\\+(k_5^2+k_6^2)(k_3^3+k_4^3-k_1^3-k_2^3)\end{matrix}\right\}\bigg/D$$

$$v_{6,y}\{k_1,k_2,k_3,k_4,k_5,k_6\} = \left\{\begin{matrix}(k_1+k_2)(k_5^3+k_6^3-k_3^3-k_4^3)\\+(k_3+k_4)(k_1^3+k_2^3-k_5^3-k_6^3)\\+(k_5+k_6)(k_3^3+k_4^3-k_1^3-k_2^3)\end{matrix}\right\}\bigg/D$$

$$D = \left\{\begin{matrix}(k_1^2+k_2^2)(k_3+k_4-k_5-k_6)\\+(k_3^2+k_4^2)(k_5+k_6-k_1-k_2)\\+(k_5^2+k_6^2)(k_1+k_2-k_3-k_4)\end{matrix}\right\} . \quad (33)$$

For a *p*-junction with any *p* to be connected to a junction with *p'* = 3 or 4 through a single soliton with wave numbers $\{k_1,k_2\}$, it must obey Eq. (28), with $g(k_1,k_2)$ given by Eq. (31). Alas, the components of the 6-velocity vector of Eq. (33) do not obey Eq. (28) for *any* of the 15 pairs of wave numbers. Hence, if a 6-junction exists, it cannot partake in the construction of an asymptotic web. Clearly, this applies only when the, *D* in Eq. (33) does not vanish. It does vanish, for example, if all three linear factors in *D* vanish, namely, the wave numbers are not generic.

A 5-velocity vector is obtained by making two of the wave numbers in Eq. (33) coincide. There are 15 ways to do this. All resulting 5-velocity vectors obey Eq. (28) only for *one* pair of wave numbers. For example, when $k_6 = k_4$ in Eq. (33), the resulting 5-velocity vector obeys Eq. (28) *only* for the pair $\{k_3,k_5\}$. Computing

$$C_{ij} = v_{5,x}\{k_1,k_2,k_3,k_4,k_5\} - (k_i+k_j)v_{5,y}\{k_1,k_2,k_3,k_4,k_5\} + (k_i^2+k_ik_j+k_j^2) , \quad (34)$$

for all $1 \leq i < j \leq 5$, one finds that $C_{35}$ vanishes, while all other 14 $C_{ij}$ do not vanish in general. Requiring that Eq. (28) hold for four additional pairs of wave numbers (i.e., allowing other connecting single solitons besides the {3,5} single soliton), one finds that the five-wave number can contain only three or two different wave numbers, reducing the junction to a 3-junction, or to a single soliton, respectively. For example, one solution is $k_1 = k_2 = k_3$.

Thus, a 5-junction can partake in the construction of an asymptotic web only through one of the solitons that emanate from it. A soliton that emerges from the junction may be excluded from the role of a connecting soliton only if long-range "memory" of the structure of the junction exists along it. However, such "memory" is impossible. Owing to the construction of solutions of Eq. (1) via Eqs. (2)-(5), all information regarding the structure of a junction decays exponentially away fro the junction along any soliton line that emerges from it. The empirical observations that asymptotically far junctions do not affect one another, hence cannot "dictate" to a given junction which of its solitons can or cannot partake in the construction of a web support this statement.

In summary, 5- and 6-junctions cannot partake in the construction of an asymptotic web. This does not exclude the possibility that, like the *X*-shaped solution, *p*-junctions with $p > 4$ may exist as single-junction solutions. Preliminary calculations, to be presented in a separate publication, indicate that, at least, 5- and 6-junctions cannot exist at all.

The preceding discussion applies only to solutions of the KP II equation, for which all the wave number may be treated as independent. Demanding that the wave numbers are generic (see Section 2.7) provides a sufficient condition for this requirement. If Eq. (17) is not obeyed by the set of *M* wave numbers, from which a solution is constructed, then the conclusions stated above may or may not apply. For example, *p*-junctions with $p > 4$ may appear in higher members of the KP II hierarchy when non-generic wave number sets are employed [40-43].

### 3.5 Asymptotic morphology
As $|t| \to \infty$, the morphology of a web does not change [24-43]. From the point of view of this paper, in which the role of junction velocities is exploited, this characteristic is a direct consequence of the fact that every junction moves at a constant velocity. The position of a *p*-junction tends to

$$\vec{r}_p(k_1, k_2, ..., k_p; t) = \vec{v}_p(k_1, k_2, ..., k_p)t + \vec{r}_0(k_1, k_2, ..., k_p) \ . \tag{35}$$

The distance between two junctions then tends to

$$\left(\vec{v}_p(k_1, k_2, ..., k_p) - \vec{v}_{p'}(k'_1, k'_2, ..., k'_{p'})\right)t + \left(\vec{r}_0(k_1, k_2, ..., k_p) - \vec{r}_0(k'_1, k'_2, ..., k'_{p'})\right) \ . \tag{36}$$

As $|t| \to \infty$, the finite initial shifts cannot modify the morphology of the web. As a result, if one chooses to plot the web against scaled coordinates,

$$\chi = \frac{x}{t} \ , \quad \eta = \frac{y}{t} \quad (|t| \to \infty) \ , \tag{37}$$

then, as $|t| \to \infty$, the structure of the web freezes and the junctions are points on a "lattice". The coordinates of the position of each junction in the $\chi$-$\eta$ plane is just its velocity vector. Examining Eq. (8), one deduces that, as $|t| \to \infty$, the width of the profile of a single-soliton decreases in the $\chi$-$\eta$ plane as $(1/|t|)$. A single soliton degenerates into a line of vanishing thickness.

Finally, as $|t| \to \infty$, the plots in the $\chi$-$\eta$ plane of the actual solutions coincide with the contour plots of Refs. [40-43], which are defined as the lines in the $\chi$-$\eta$ plane, along which the logarithm of each term in the function $f(t,x,y)$ of Eq. (3) attains its local maximum:

$$\lim_{t \to \pm\infty} \left( \max_{1 \leq i_1 < ... < i_N \leq M} \left\{ \log \left[ \xi_M(i_1, ..., i_N) \left( \prod_{1 \leq j < l \leq N} (k_{i_l} - k_{i_j}) \right) \right] + \sum_{j=1}^{N} \theta_{i_j}(t, x = t\chi, y = t\eta) \right\} \right) \ . \tag{38}$$

**4. Numerical examples of asymptotic webs: (M,N) solutions, $M \geq 4$, $N \geq 2$**

The characteristics of the structure of solutions presented here coincide with the conclusions of Refs. [40-43], but can be also demonstrated rigorously through an analysis of the solutions employing a "riding on junctions" procedure. For the sake of clarity of presentation, such demonstrations are deferred to a future publication. For example, the participation of 4-junctions in a web, is possible only if at least one Plücker coordinate in Eqs. (2)-(5) vanishes. In higher members of the KP II hierarchy (not discussed here), $p$-junctions with $p > 4$ can partake in the construction of webs, provided the Plücker coordinates obey certain constraints and the wave numbers are not-generic [40-43]. However, the fact that the wave number set is not generic does not necessarily imply degeneracy in the solution. For example, in some of the numerical examples, the wave numbers constitute a Fibonacci sequence, hence, are not generic. Still, the asymptotic web-structure displays the same universal characteristics. Finally, in all the figures, pairs, $\{i,j\}$, of integers near a soliton represent the wave numbers, $\{k_i, k_j\}$, from which it is constructed.

### 4.1 Only 3-junctions when all Plücker coordinates ≠ 0
Figs. 4 and 5 show the $t \gg 0$ webs for, respectively, (4,2) and (5,2) solutions, in which none of the $\xi$- coefficients vanishes. The $t \ll 0$ webs (not shown) also display 3-junctions only.

### 4.2 4-junctions emerge when at least one Plücker coordinate = 0
#### 4.2.a (4,2) solution
The *X*-solution (see Fig. 2) is a well-known example of a single-4-junction solution [5,23-33]. It is a (4,2) solution, obtained when *two* of the $\xi_4$ coefficients vanish (see Eq. (13)). The junction persists as $t \to \pm\infty$. However, a 4-junction may partake in the construction of an asymptotic web also if only *one* $\xi_4$ coefficient is made to vanish. In this case, a 4-junction may appear either as $t \to +\infty$ or as $t \to -\infty$. Where it appears depends on the numerical values of the wave numbers. In the other limit, the web contains only 3-junctions. The asymptotic web of a solution with $\xi_4(1,2) = 0$ is shown in Figs. 6 and 7 for, respectively, $t \gg 0$ and $t \ll 0$.

#### 4.2.b (5,2) solution
Figs. 8 and 9 present the structure of the, respectively, $t \gg 0$ and $t \ll 0$ asymptotic web of a (5,2) solution that contains 3- and 4-junctions. The choice, $\xi_5(3,4) = 0$, is responsible for the generation of a 4-junction. Different 4-junctions emerge as $t \to +\infty$ and as $t \to -\infty$.

#### 4.2.c (6,3) solution
Fig 10 shows the $t \gg 0$ asymptotic web of a (6,3) solution, in which $\xi_6(1,2,3) = \xi_6(4,5,6) = 0$. Up to space inversion, $\{x,y\} \to \{-x, -y\}$, this solution has the same structure for $t > 0$ and for $t < 0$ [30, 31]. Hence, the same 3- and 4-junctions emerge in the asymptotic web also for $t \ll 0$.

### 4.3 Junction "lattice"

Figs. 11 and 12 each show two superimposed plots for a (5,2) solution with an asymptotic web that contains only 3-junctions. The plots are plotted against the scaled coordinates $\chi$ and $\eta$ of Eq. (37). The first plot is the solution. The second plot is the junction "lattice", where the coordinates of each junction are the components of its 3-velocity vector, computed by Eq. (12). In Fig. 11, time is short enough, so that the thickness of single-soliton lines in the $\chi$-$\eta$ plane is substantial, and the asymptotic "lattice" points miss the actual junctions. In Fig. 12, time is long enough, so that the shrinking of the thickness of single solitons (towards lines of vanishing thickness) is visible, and, as expected, the "lattice" points coincide with the locations of the junctions in the asymptotic web.

Acknowledgment Y.Z. wishes to thank Y. Kodama for very instructive discussions.

## Appendix A. Derivation of Plücker relations

A simple way to obtain the Plücker relations for the $\xi$-coefficients in an $(M,N)$ solution (see Eq. (3)) is through the observation that these coefficients may be represented as the determinants of $N \times N$ minors of an $N \times M$ matrix.

<u>(4,2) solution</u> The matrix is

$$A = \begin{pmatrix} a_{11} & a_{12} & a_{13} & a_{14} \\ a_{21} & a_{22} & a_{23} & a_{24} \end{pmatrix} . \tag{A.1}$$

The $\xi_4$-coefficients are defined by

$$\xi_4(i,j) = \begin{vmatrix} a_{1,i} & a_{1,j} \\ a_{2,i} & a_{2,j} \end{vmatrix} , \quad (i<j) . \tag{A.2}$$

With eight matrix elements and six $\xi$-coefficients, this definition yields one constraint:

$$\xi_4(1,2)\xi_4(3,4) - \xi_4(1,3)\xi_4(2,4) + \xi_4(1,4)\xi_4(2,3) = 0 . \tag{A.3}$$

<u>(5,2) solution</u> The matrix is

$$A = \begin{pmatrix} a_{11} & a_{12} & a_{13} & a_{14} & a_{15} \\ a_{21} & a_{22} & a_{23} & a_{24} & a_{25} \end{pmatrix} . \tag{A.4}$$

With ten matrix elements and 10 $\xi$-coefficients, this definition yields five constraints:

$$\begin{aligned}
\xi_5(1,2)\xi_5(3,4) - \xi_5(1,3)\xi_4(2,4) + \xi_5(1,4)\xi_4(2,3) &= 0 \\
\xi_5(1,2)\xi_5(3,5) - \xi_5(1,3)\xi_4(2,5) + \xi_5(1,5)\xi_4(2,3) &= 0 \\
\xi_5(1,2)\xi_5(4,5) - \xi_5(1,4)\xi_4(2,5) + \xi_5(1,5)\xi_4(2,4) &= 0 \\
\xi_5(1,3)\xi_5(4,5) - \xi_5(1,4)\xi_4(3,5) + \xi_5(1,5)\xi_4(3,4) &= 0 \\
\xi_5(2,3)\xi_5(4,5) - \xi_5(2,4)\xi_4(3,5) + \xi_5(2,5)\xi_4(3,4) &= 0
\end{aligned} . \tag{A.5}$$

<u>(6,2) solution</u> The matrix is

$$A = \begin{pmatrix} a_{11} & a_{12} & a_{13} & a_{14} & a_{15} & a_{16} \\ a_{21} & a_{22} & a_{23} & a_{24} & a_{25} & a_{26} \end{pmatrix} . \tag{A.6}$$

With twelve matrix elements and 15 $\xi$-coefficients, this definition yields twelve constraints:

$$\xi_5(1,2)\xi_5(3,4) - \xi_5(1,3)\xi_4(2,4) + \xi_5(1,4)\xi_4(2,3) = 0$$
$$\xi_5(1,2)\xi_5(3,5) - \xi_5(1,3)\xi_4(2,5) + \xi_5(1,5)\xi_4(2,3) = 0$$
$$\xi_5(1,2)\xi_5(3,6) - \xi_5(1,3)\xi_4(2,6) + \xi_5(1,6)\xi_4(2,3) = 0$$
$$\xi_5(1,2)\xi_5(4,5) - \xi_5(1,4)\xi_4(2,5) + \xi_5(1,5)\xi_4(2,4) = 0$$
$$\xi_5(1,2)\xi_5(4,6) - \xi_5(1,4)\xi_4(2,6) + \xi_5(1,6)\xi_4(2,4) = 0$$
$$\xi_5(1,3)\xi_5(4,5) - \xi_5(1,4)\xi_4(3,5) + \xi_5(1,5)\xi_4(3,4) = 0$$
$$\xi_5(1,3)\xi_5(4,6) - \xi_5(1,4)\xi_4(3,6) + \xi_5(1,6)\xi_4(3,4) = 0 \quad . \tag{A.7}$$
$$\xi_5(1,3)\xi_5(5,6) - \xi_5(1,5)\xi_4(3,6) + \xi_5(1,6)\xi_4(3,5) = 0$$
$$\xi_5(2,3)\xi_5(4,5) - \xi_5(2,4)\xi_4(3,5) + \xi_5(2,5)\xi_4(3,4) = 0$$
$$\xi_5(2,3)\xi_5(4,6) - \xi_5(2,4)\xi_4(3,6) + \xi_5(2,6)\xi_4(3,4) = 0$$
$$\xi_5(2,4)\xi_5(5,6) - \xi_5(2,5)\xi_4(4,6) + \xi_5(2,6)\xi_4(4,5) = 0$$
$$\xi_5(3,4)\xi_5(5,6) - \xi_5(3,5)\xi_4(4,6) + \xi_5(3,6)\xi_4(4,5) = 0$$

(6,3) solution  The matrix is

$$A = \begin{pmatrix} a_{11} & a_{12} & a_{13} & a_{14} & a_{15} & a_{16} \\ a_{21} & a_{22} & a_{23} & a_{24} & a_{25} & a_{26} \\ a_{31} & a_{32} & a_{33} & a_{34} & a_{35} & a_{36} \end{pmatrix} \quad . \tag{A.8}$$

With eighteen matrix elements and 20 $\xi$-coefficients, this definition yields the constraints:

$$\xi_6(i,j,k)\xi_6(i,l,m) - \xi_6(i,j,l)\xi_6(i,k,m) + \xi_6(i,j,m)\xi_6(i,k,l) = 0 \quad , \tag{A.9}$$

with all permutations of the indices.

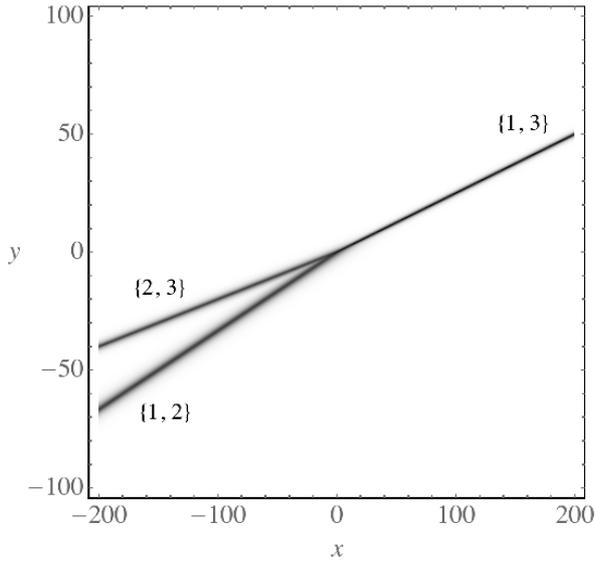

Fig. 1 *Y*-shaped (3,1)-soliton solution.
$\xi_3(1) = \xi_3(2) = \xi_3(3) = 1$; $k_1 = 1$, $k_2 = 2$, $k_3 = 3$; $t = 0$.

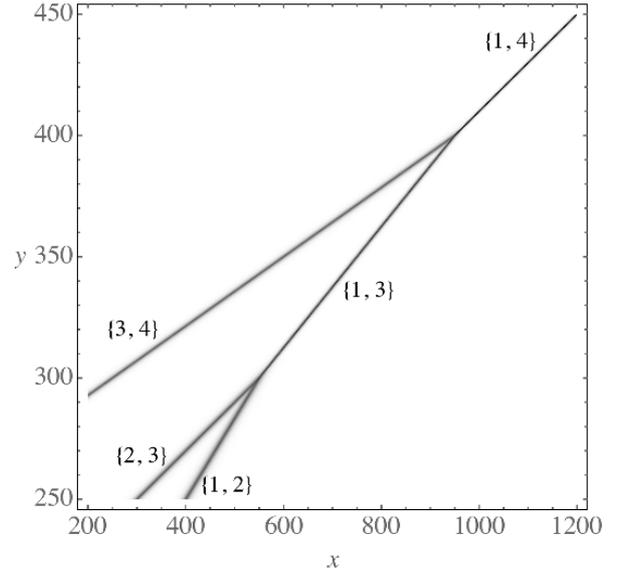

Fig. 3 (4,1)-soliton solution.
$\xi_4(1) = \xi_4(2) = \xi_4(3) = \xi_4(4) = 1$;
$k_1 = 1$, $k_2 = 2$, $k_3 = 3$, $k_4 = 4$; at $t = 50$.

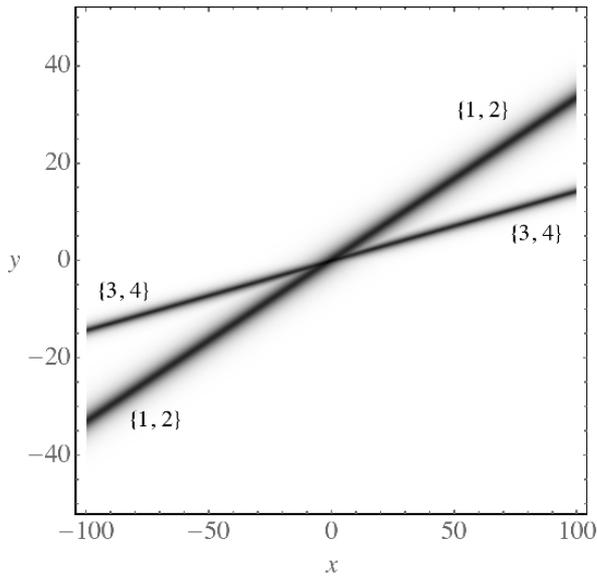

Fig. 2 *X*-shaped (4,2)-soliton solution.
$\xi_4(1,2) = \xi_4(3,4) = 0$, $\xi_4(1,3) = \xi_4(1,4) = \xi_4(2,3) = \xi_4(2,4) = 1$;
$k_1 = 1$, $k_2 = 2$, $k_3 = 3$, $k_4 = 4$; $t = 0$.

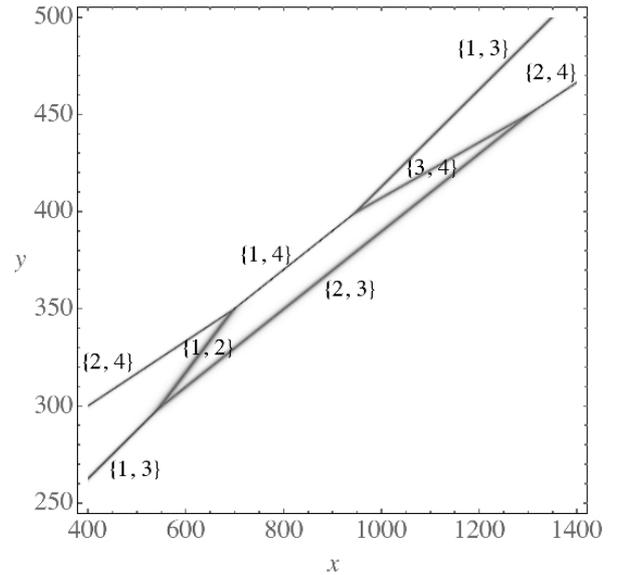

Fig. 4 (4,2)-soliton solution.
$\xi_4(i,j) = 2\dfrac{(j-i)}{(i+2)(j+2)} \neq 0$;
$k_1 = 1$, $k_2 = 2$, $k_3 = 3$, $k_4 = 4$; $t = 50$.

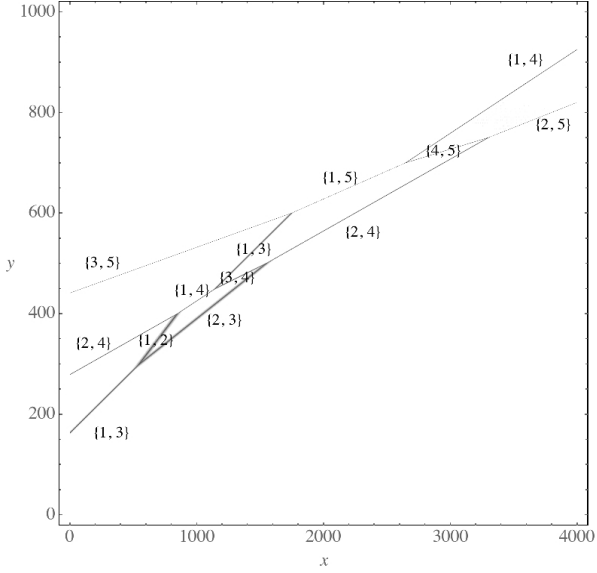
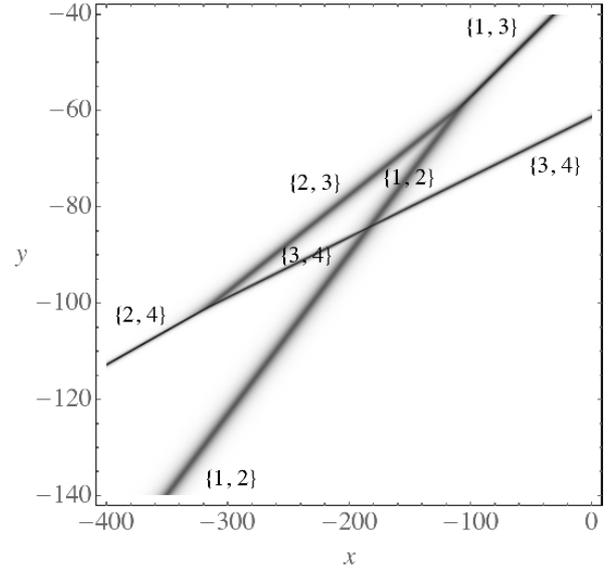

Fig. 5 (5,2)-soliton solution

$$\xi_5(i,j) = 2\frac{(j-i)}{(i+2)(j+2)} \neq 0 \;;$$

$k_1 = .1$, $k_2 = .2$, $k_3 = .3$, $k_4 = .5$, $k_5 = .8$; $t = 10{,}000$.

Fig. 7 (4,2)-soliton solution. Parameters as in Fig. 6; $t = -10$.

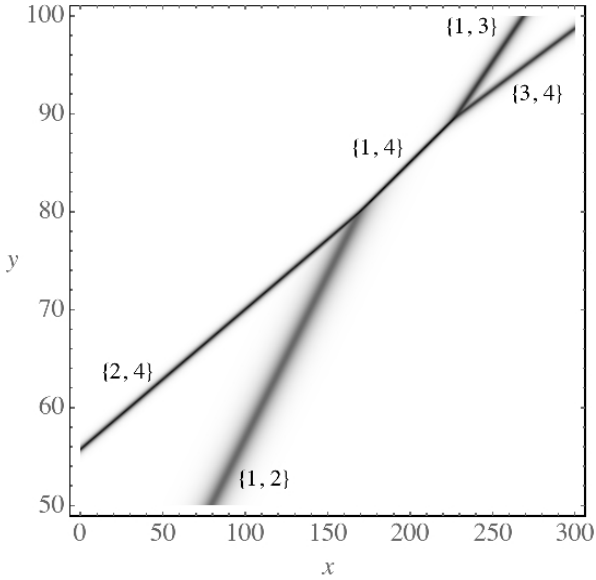
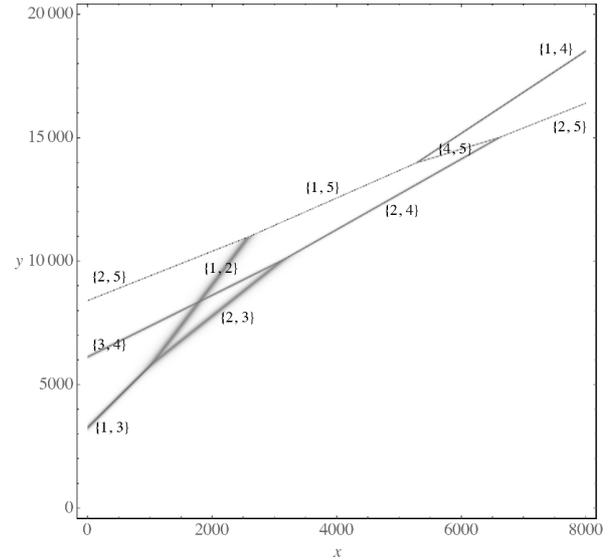

Fig. 6 (4,2)-soliton solution. $\xi_4(1,2) = 0$, $\xi_4(1,3) = \frac{4}{15}$, $\xi_4(1,4) = \frac{1}{3}$, $\xi_4(2,3) = \frac{4}{15}$, $\xi_4(2,4) = \frac{1}{3}$, $\xi_4(3,4) = \frac{1}{15}$;

$k_1 = 1$, $k_2 = 2$, $k_3 = 3$, $k_4 = 5$; $t = 10$.

Fig. 8 (5,2) solution. $\xi_5(3,4) = 0$, $\xi_5(1,2) = \frac{1}{6}$, $\xi_5(1,3) = \xi_5(1,4) = \frac{4}{15}$, $\xi_5(1,5) = \frac{8}{21}$, $\xi_5(2,3) = \xi_5(2,4) = \frac{1}{10}$, $\xi_5(2,5) = \frac{3}{14}$, $\xi_5(3,5) = \xi_5(4,5) = \frac{4}{35}$;

$k_1 = .1$, $k_2 = .2$, $k_3 = .3$, $k_4 = .5$, $k_5 = .8$; $t = 10{,}000$.

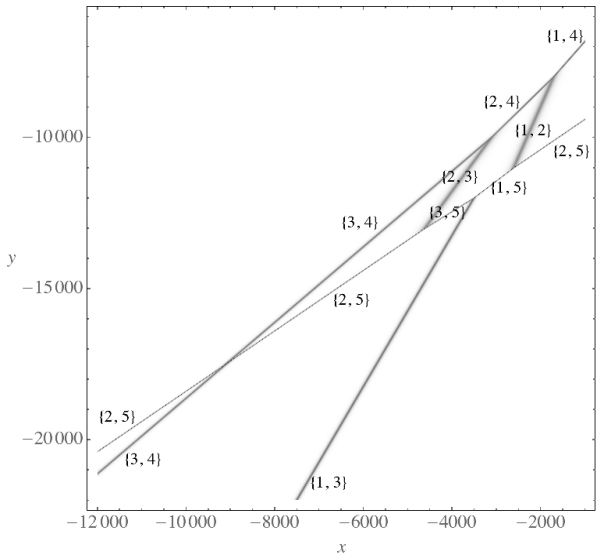

Fig. 9 (5,2) solution. Parameters as in Fig. 8. $t = -10{,}000$.

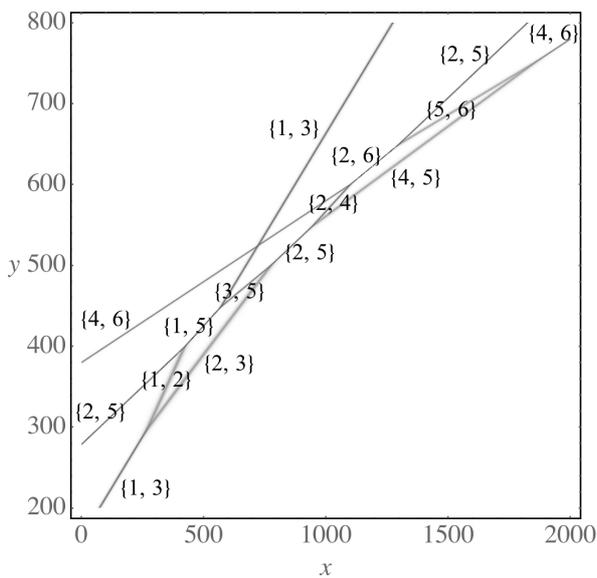

Fig. 10 (6,3) solution. $\xi_6(1,2,3) = \xi_6(4,5,6) = 0$, $\xi_6(1,4,6) = 2$, all remaining $\xi_6(i,j,l) = 1$; $k_m = m/2$; $t = 100$.

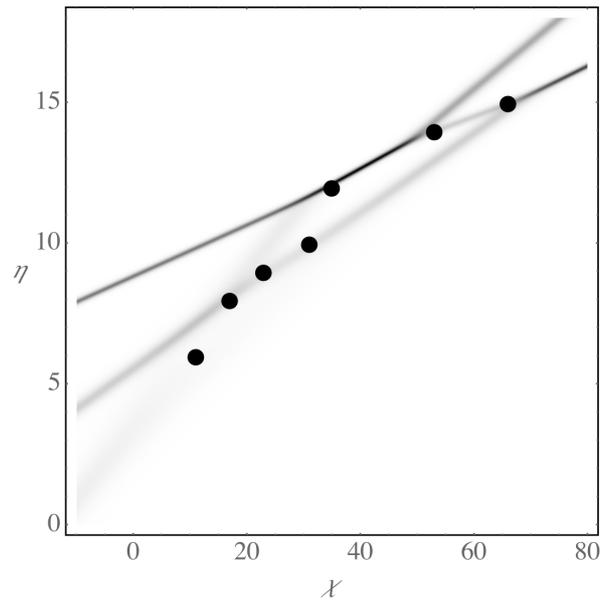

Fig. 11 (5,2) solution $\xi_5(i,j) = 2\dfrac{(j-i)}{(i+2)(j+2)} \neq 0$; $k_1 = 1$, $k_2 = 2$, $k_3 = 3$, $k_4 = 5$, $k_5 = 8$, vs. scaled coordinates (Eq. (37)) together with junction lattice. Lattice-point coordinates are components of 3-velocities of junctions (Eq. (12)). $t = 0.4$.

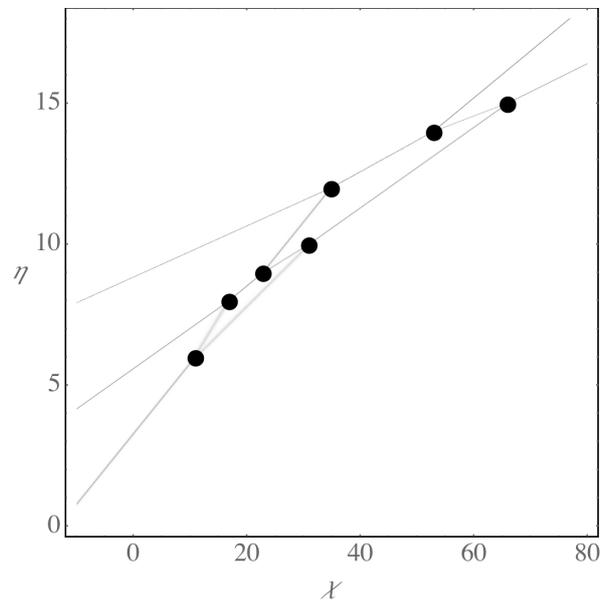

Fig. 12 (5,2) solution (parameters as in Fig.11), vs. scaled coordinates (Eq. (37)) together with junction lattice. $t = 10$.